\documentclass[10pt,reqno,twoside,paper=letter]{amsart}

\usepackage[letterpaper,left=20mm,right=14mm,layoutvoffset=10mm]{geometry}
\usepackage{graphicx}
\usepackage{epstopdf}

\usepackage[bf,small]{titlesec}
\titlelabel{\thetitle.\,\,\,}
\usepackage{amssymb,amsfonts}
\usepackage{amsmath}
\usepackage{float} % para usar [H]
\usepackage{multicol}
\usepackage{fancyhdr}
\usepackage[latin1]{inputenc}				% For DOS systems
\usepackage[comma,longnamesfirst,sectionbib]{natbib}
\usepackage{enumerate}
\usepackage[symbol*]{footmisc}

\newcommand{\inforevista}{\scriptsize  Rev. Acad. Colomb. Cienc. Ex. Fis. Nat. 42(XXX):XX--XX, enero-marzo de 2019}
\begin{document}
%\printinunitsof{mm}\prntlen{\textwidth} % check the textwidth length.
%
%% these are fancyhdr related
%
%\pagenumbering{arabic}
%\pagenumbering{gobble}
%\setcounter{page}{35}
\fancypagestyle{plain}{%
\fancyhf{} % clear all header and footer fields
\fancyfoot[R]{\thepage} %
\fancyhead[L]{\inforevista}
\renewcommand{\headrulewidth}{0pt}
\renewcommand{\footrulewidth}{0pt}}
%%%
\thispagestyle{empty} %for first page
%%%
\pagestyle{empty}
\fancyhead{} % clear all header fields
\renewcommand{\headrulewidth}{0.0pt} %remove a decorative line from header
\fancyhead[LO]{\inforevista}
\fancyhead[RO]{\scriptsize Orbital dynamics in galactic models: NGC 3726, 3877 and 4010}
\fancyhead[LE]{\scriptsize F. L. Dubeibe, S. M. Mart\'inez-Sicach\'a, G. A. Gonz\'alez}
\fancyhead[RE]{\inforevista}
\fancyfoot{} % clear all footer fields
\fancyfoot[LE]{\thepage}
\fancyfoot[RO]{\thepage}

\begin{flushright}
\rule[-0.5ex]{0.5ex}{3.0ex} {\large Physical Sciences}
\end{flushright}
\vspace*{0.3cm}

\begin{center}
{\LARGE \textbf{Orbital dynamics in realistic galaxy models: NGC 3726, NGC 3877 and NGC 4010\\}}
\end{center}

\vspace{3mm}
\begin{center}
\textbf{\small Fredy L. Dubeibe${}^{1}$, Sandra M. Mart\'inez-Sicach\'a${}^{2}$, Guillermo A. Gonz\'alez${}^{2}\footnotemark[1]$}
\vspace{3mm}

{\scriptsize
${}^{1}$Grupo de Investigaci\'on Cavendish, Facultad de Ciencias Humanas y de la Educaci\'on, Universidad de los Llanos, Villavicencio, Colombia\\
${}^{2}$Grupo de Investigaci\'on en Relatividad y Gravitaci\'on, Escuela de F\'isica, Universidad Industrial de Santander, A.A. 678, Bucaramanga, Colombia\\
 }
\end{center}
\footnotetext[1]{Correspondencia: G. A. Gonz\'alez, guillermo.gonzalez@saber.uis.edu.co, Recibido: 5 de octubre de 2018; Aceptado: 14 de febrero de 2019.}

\begin{Small}
\vspace{3.0mm}
\rule{\textwidth}{0.4pt}

\begin{center}
\begin{minipage}{14cm}
\vspace{3mm}

\textbf{Abstract}
\vspace{3mm}

{In the present paper, using a generalization of the Miyamoto and Nagai potential we adjusted the observed rotation curves of three specific spiral galaxies to the analytical circular velocities. The observational data have been taken from a 21 cm-line synthesis imaging survey using the Westerbork Synthesis Radio Telescope, for three particular galaxies in the Ursa Major cluster: NGC 3726, NGC 3877 and NGC 4010. Accordingly, the dynamics of the system is analyzed in terms of the Poincar\'e sections method, finding that for larger values of the angular momentum of the test particle or lower values its total energy the dynamics is mainly regular, while on the opposite cases, the dynamics is mainly chaotic. Our toy model opens the possibility to find chaotic bounded orbits for stars in those particular galaxies.
\\[1mm]

\textbf{Key words:}  Stellar dynamics; Galaxies: kinematics and dynamics; Nonlinear dynamics and chaos.}

\vspace{3mm}

\textbf{Din\'amica orbital en modelos realistas de gal\'axias: NGC 3726, NGC 3877 y NGC 4010}
\vspace{3mm}

%\flushleft{}
\textbf{Resumen}
\vspace{3mm}

{En el presente trabajo, utilizando una generalizaci\'on del potencial de Miyamoto-Nagai, se ajustan las curvas de rotaci\'on observadas de tres galaxias espirales a las velocidades circulares anal\'iticas. Los datos observacionales se tomaron de un conjunto de im\'agenes de l\'inea de 21 cent\'imetros (o l\'inea HI) 
obtenidos con el Westerbork Synthesis Radio Telescope (WSRT), para tres galaxias particulares en el grupo de la Ursa Major: NGC 3726, NGC 3877 y NGC 4010. Seguidamente, se analiza la din\'amica del sistema en t\'erminos del m\'etodo de secciones de Poincar\'e, encontrando que para valores grandes del momento angular de la part\'icula de prueba o valores bajos su energ\'ia total, la din\'amica es principalmente regular, mientras que en los casos opuestos, la din\'amica es principalmente ca\'otica. Nuestro modelo abre la posibilidad de encontrar \'orbitas ca\'oticas acotadas para estrellas presentes en esas galaxias particulares.
\\[1mm]

\textbf{Palabras clave:} Din\'amica estelar, Galaxias: cinem\'atica y din\'amica, Din\'amica no lineal y caos.}
\end{minipage}
\end{center}

%\end{adjustwidth}
\rule{\textwidth}{0.4pt}
\end{Small}

\begin{small}
\columnsep 0.5 cm
\begin{multicols}{2}
%\sloppy

\setlength{\parskip}{.3cm}

\section*{Introduction}

Since the seminal paper by \cite{Miyamoto1975}, the literature on three-dimensional analytical models for the gravitational field of different types of galaxies has grown considerably. In this respect, particular attention deserve the models proposed by \cite{Jaffe1983} and \cite{Hernquist1990}, who derived analytical models which closely approximate the light distribution for spherical and elliptical galaxies, respectively. A few years later, \cite{Long1992} presented an analytical potential for barred galaxies that reduces to the Miyamoto-Nagai disk by an appropriate setting of the free parameters, while \cite{Dehnen1993}  generalized the Jaffe and Hernquist models by means of a family of density profiles with different central slopes. More recently, \cite{Vogt2005} derived an analytical expression for the gravitational field of galaxies, based on the multipolar expansion up to the quadrupole term. Using a different approach, \cite{Gonzalez2010} obtained a family of finite thin-discs models for four galaxies in the Ursa major cluster in which the circular velocities were adjusted to fit the observed rotation curves.

One advantage of an analytical galaxy model is the possibility to study the dynamics (regular or chaotic) of orbits. This can be considered one of the standing problems in galactic dynamics because it could allow us to understand the formation and evolution of galaxies \citep{Contopoulos1979}, as shown by the pioneer simulations of \cite{Lindblad}. Despite the fact that early papers on this topic studied only regular orbits in the meridional plane  \citep{Martinet1975,Manabe1979,Greiner1987,Lees1992}, soon after, the existence of chaos on the orbital motion started to be considered by \cite{Caranicolas1996} and \cite{Caranicolas2003}. In the majority of cases all these studies focused on the distinction between regular and chaotic orbits \citep{Manos2011,Bountis2012,Manos2013} or the influence of the galaxy components (nucleus, bulge, disk, halo) on the character of orbits, see e.g. \citep{Zotos2012,Zotos2013,Zotos2014}. Notwithstanding the evidence that both chaotic and regular motions are possible in many axisymmetric potentials, recent studies on generalized axisymmetric potentials suggest that a third integral of motion seems to exist for energy values closer to the escape energy \citep{Dubeibe2018,Zotos2018}. Hence, such apparent ambiguity might only be solved by performing systematic studies of each particular model. 

In this paper, we are interested in meridional motions of free test particles (stars) in presence of analytical realistic galaxy models. Our models possess axial symmetry, which is a good approximation given the morphology of galaxies that are mainly approximate figures of revolution. Additionally, the galaxy components were not added one by one, instead of this, we derived a generalized Miyamoto-Nagai model that can be adjusted very accurately to fit the observed rotation curve and hence it is assumed that all (or most of) the components are taken into account. The determination of the specific values of the coefficients of the series expansion let us calculate the corresponding surface densities and all the kinematic quantities characterizing the particular galaxy models. Unlike the models derived by \cite{Gonzalez2010}, which exhibit instabilities to small vertical perturbations (see {\it e.g.} the cases of NGC 3877 and NGC 4010), our models satisfy the stability conditions for radial and vertical perturbations. On the other hand, the dynamics of the orbits is studied through the Poincar\'e surfaces of section, showing that the orbital motion exhibits a strong dependence on the angular momentum and energy of the test particles (stars).

The paper is organized as follows: in the first section, we derive the generalized Miyamoto-Nagai model. Next, from the new potential the explicit expressions for the physical quantities of interest are determined. In the second section we adjust the observed rotation curves of three specific spiral galaxies (NGC 3726, NGC 3877 and NGC 4010) to the analytical circular velocities derived with our model. Then, the mass-density profiles are calculated, along with the vertical and epicyclic frequencies, showing that our model not only is well-behaved but also satisfy the stability conditions. A dynamical analysis in terms of the Poincar\'e surfaces of section is performed in the third section. Finally, in the fourth section, we summarize our main conclusions. 

\section*{Generalized Miyamoto-Nagai model}
\label{sec:1}

Let us start considering the axially symmetric Laplace's equation in spherical coordinates
\begin{equation}\label{eq1}
\nabla^2\Phi(r,\theta)=\frac{1}{r^2}\frac{\partial}{\partial r}\left(r^2 \frac{\partial \Phi}{\partial r}\right)+\frac{1}{r^2 \sin\theta}\frac{\partial}{\partial \theta}\left(\sin\theta \frac{\partial \Phi}{\partial \theta}\right)=0,
\end{equation}
whose general solution reads as 
\begin{equation}\label{eq2}
\Phi(r,\theta)=\sum_{l=0}^{\infty}\left(A_{l}\, r^{l}- B_{l}\,  r^{-(l+1)}\right) P_{l}(\cos\theta),
\end{equation}
where $A_{l}$ and $B_{l}$ are constants to be determined, $P_{l}$ are the Legendre polynomials, and the notation $(r,\theta,\phi)$ means (radial, polar, azimuthal) coordinates, respectively.

Since $\Phi(r,\theta)$ denotes the gravitational potential of an axisymmetric finite distribution of mass, the boundary condition $\lim_{r\to\infty} \Phi(r,\theta)=0$ must be satisfied, thus the solution \eqref{eq2} takes the form 
\begin{equation}\label{eq3}
\Phi(r,\theta)=-\sum_{l=0}^{\infty}\frac{B_{l}  P_{l}(\cos\theta)}{r^{l+1}}.
\end{equation}
Following \cite{Vogt2005}, in order to obtain a generalized Miyamoto-Nagai model and for the sake of simplicity, we shall consider terms up to $l=3$ in \eqref{eq3}, therefore, transforming to cylindrical coordinates $(R,z)$ by means of the relations 
\begin{equation}
\cos\theta=z/r \quad {\rm and} \quad r=\sqrt{R^{2}+z^2}, 
\end{equation}
and applying the additional transformation \citep{Satoh1980},
\begin{equation}
z\rightarrow z^{*}=a+\sqrt{z^{2}+b^2}, 
\end{equation}
with $a$ and $b$ two arbitrary parameters, the generalized potential takes the form\footnote{It should be noted that setting $B_0 = GM, B_1 = B_2 = B_3 = 0$ in \eqref{eq4}, we get the well-known Miyamoto-Nagai Potential \citep{Miyamoto1975}.}
\begin{eqnarray}\label{eq4}
\Phi(R,z)&=&-\frac{B_0}{\sqrt{{R^2+z^{*}}^2}} -\frac{B_1\,
   z^{*}}{\left(R^2 + {z^{*}}^2\right)^{3/2}}\nonumber\\ 
   &+&\frac{B_2 \left(R^2 - 2 {z^{*}}^2\right)}{2
   \left(R^2+ {z^{*}}^2\right)^{5/2}} +\frac{B_3 \left(3 R^{2} z^{*}- 2 {z^{*}}^{3} \right)}{ 2 \left(R^2 + {z^{*}}^2\right)^{7/2}}.
\end{eqnarray}

Once the potential has been specified, the mass-density distribution $\Sigma$ can be calculated directly from Poisson equation,
\begin{equation}\label{eq5}
\Sigma= \frac{1}{4\pi G} \left(\frac{\partial^{2}\Phi}{\partial R^2}+\frac{1}{R}\frac{\partial \Phi}{\partial R}+\frac{\partial^{2}\Phi}{\partial z^2} \right),
\end{equation}
while the circular velocity $v$ of particles in the galactic plane, the epicyclic frequency $k$, and the vertical frequency $\nu$ of small oscillations about the equilibrium circular orbit, can be obtained from the following expressions evaluated at $z=0$ \citep{Binney2011} 
\begin{eqnarray}
 v^{2} &=& R \frac{\partial \Phi}{\partial R}, \label{eq6}\\
 k^{2} &=& \frac{\partial^2 \Phi}{\partial R^2}+\frac{3}{R}  \frac{\partial \Phi}{\partial R}, \label{eqEF}\\
 \nu^{2} &=& \frac{\partial^2 \Phi}{\partial z^2}. \label{eqVF}
 \end{eqnarray}
From (\ref{eq6}-\ref{eqVF}),  it is important to emphasize that a feasible model must satisfy the constraints set by the conditions $v^{2}\ge 0$, $k^2 \ge 0$, and $\nu^2 \ge 0$, where the last two inequalities are understood as stability conditions \citep{Vogt2005}.

As is evident from the preceding paragraphs, the galactic models and its associated physical quantities are uniquely determined by the set of constants $a, b, B_0, B_1, B_2$, and $B_3$, which (taking a pragmatic approach) can be estimated from the observational data of the corresponding rotation curves, as we will discuss in detail in the next section. 

\section*{Rotation curves fitting}\label{sec:2}

The observational data were taken from \cite{Verheijen2001} for three specific galaxies in the Ursa Major cluster: NGC 3726, NGC 3877 and NGC 4010. Following the procedure outlined in \cite{Gonzalez2010}, we take the galaxy radius $R_{d}$ as the given by the largest tabulated value of the data. Thus,  introducing dimensionless variables $\tilde{R}=R/R_{d}, \tilde{z}=z/R_{d}, \tilde{a}=a/R_{d}, \tilde{b}=b/R_{d}$ and setting $\tilde{B_{0}}=B_{0}/R_{d}, \tilde{B_{1}}=B_{1}/R_{d}^{2}, \tilde{B_{2}}=B_{2}/R_{d}^{3}$, and $\tilde{B_{3}}=B_{3}/R_{d}^4$, the nonlinear least square curve fitting method allows us to calculate the numerical values of the parameters for each particular galaxy. The resulting values of $\tilde{a}, \tilde{b}, \tilde{B_0}, \tilde{B_1}, \tilde{B_2}$, and $\tilde{B_3}$, for the three galaxies under consideration, are given in Table 1.

\begin{center}
\begin{tabular}{@{}llll@{}}
\hline
  & NGC 3726 & NGC 3877 & NGC 4010\\
\hline
$\tilde{a}$ &$0.6773$ & $0.8491$ & $1.143$\\
$\tilde{b}$ &$-1.045\times10^{-6}$ & $-2.929\times10^{-7}$ & $-9.568\times10^{-6}$\\
$\tilde{B_{0}}$ &$-7.183\times10^{4}$ & $-9.859\times10^4$ & $-3.146\times10^4$\\
$\tilde{B_{1}}$ &$1.342\times10^5$ & $1.820\times10^5$ & $4.735\times10^4$\\
$\tilde{B_{2}}$ & $-8.337\times 10^4$ & $-1.098\times10^5$&$1.464\times10^4$\\
$\tilde{B_{3}}$ & $2.616\times10^4$ & $3.674\times10^4$&$-7.815\times10^3$\\
\hline
\end{tabular}\label{tab:2.1}\\
\end{center}
{\bf Table 1.} Parameters for each particular galaxy model.

\begin{figure}[H]
\begin{center}
\includegraphics[width=0.459\textwidth]{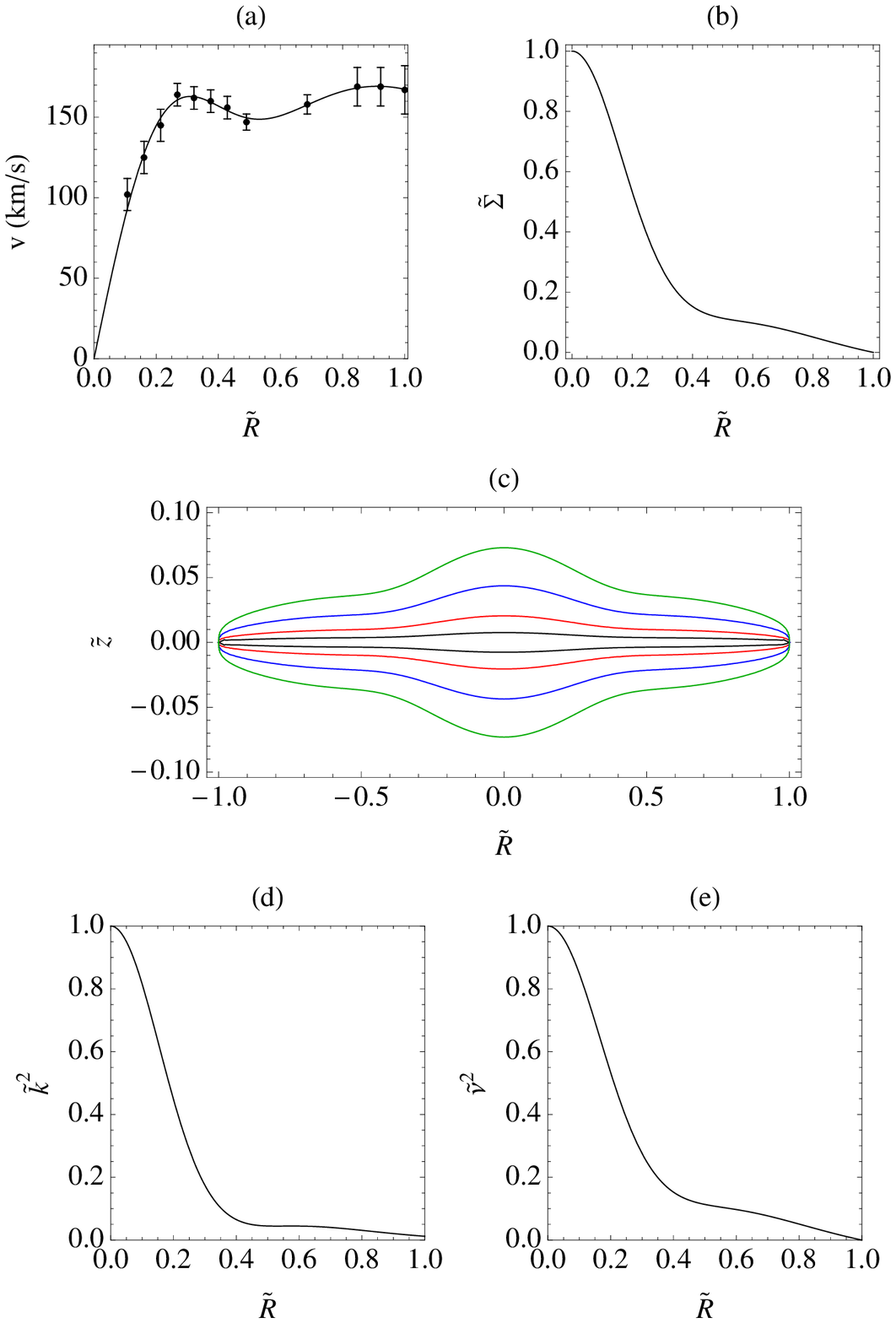}
\label{fig1}
\end{center}
\end{figure}
\vspace{-8mm}
\begin{small}
{\bf Figure 1.} Model fitted to the rotation curve of NGC 3726 using the parameters given in the first column of Table 1. (a) The solid curve indicates the rotation velocity calculated from \eqref{eq6} while the error bars denote the velocity dispersions of the observational data. (b) Normalized mass-density distribution $\tilde{\Sigma}$ at $z=0$, calculated from \eqref{eq5}. (c) Constant-density curves of equation \eqref{eq5} in the meridional plane. (d)  Epicyclic frequency \eqref{eqEF} evaluated on $z = 0$. (e) Vertical frequency \eqref{eqVF} evaluated on $z = 0$. 
\end{small}

\begin{figure}[H]
\begin{center}
\includegraphics[width=0.459\textwidth]{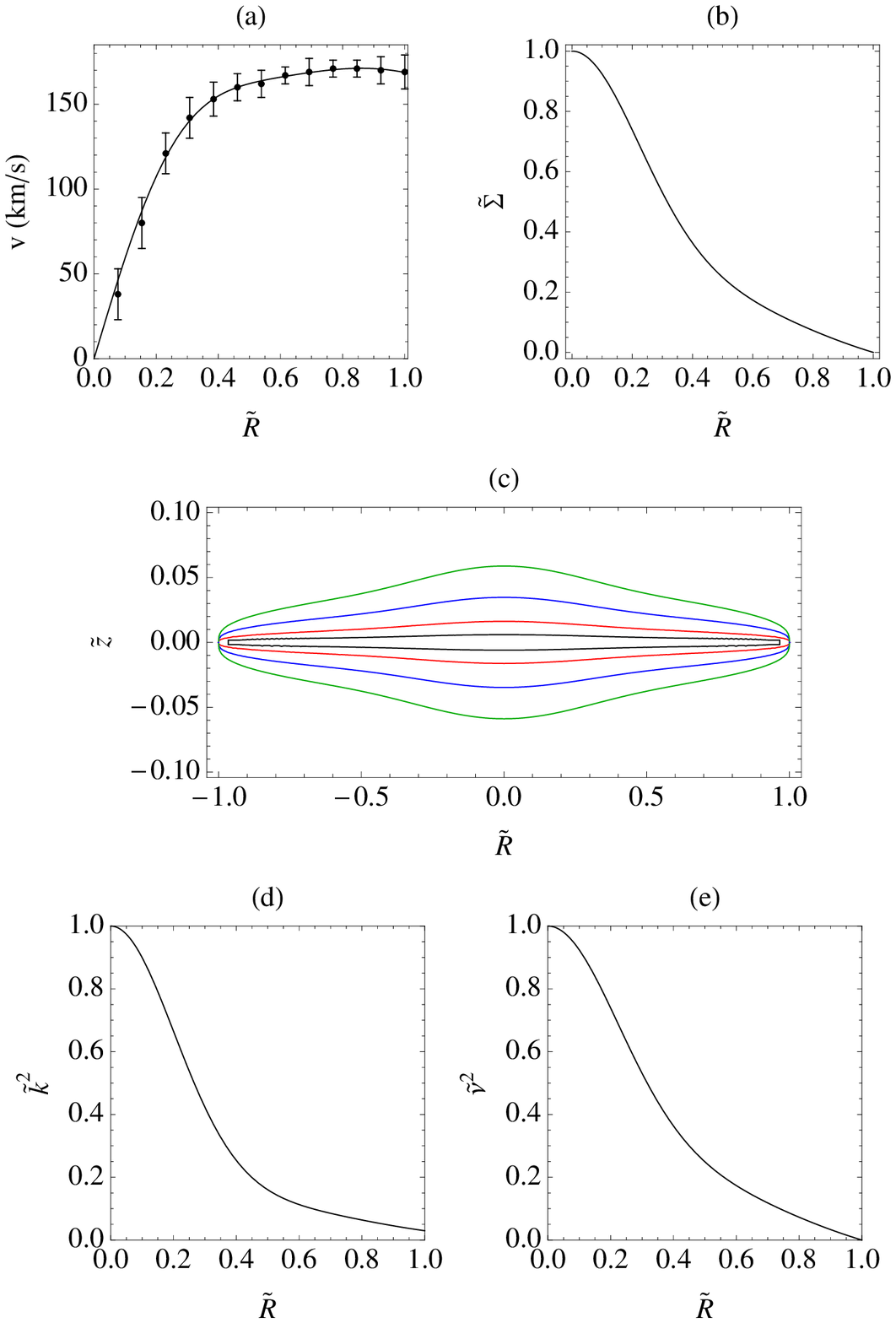}
\label{fig2}
\end{center}
\end{figure}
\vspace{-8mm}
\begin{small}
{\bf Figure 2.} Model fitted to the rotation curve of NGC 3877 using the parameters given in the second column of Table 1. (a) The solid curve indicates the rotation velocity calculated from \eqref{eq6} while the error bars denote the velocity dispersions of the observational data. (b) Normalized mass-density distribution $\tilde{\Sigma}$ at $z=0$, calculated from \eqref{eq5}. (c) Constant-density curves of equation \eqref{eq5} in the meridional plane. (d)  Epicyclic frequency \eqref{eqEF} evaluated on $z = 0$. (e) Vertical frequency \eqref{eqVF} evaluated on $z = 0$. 
\end{small}

In panels (a) of Figures 1, 2, and 3, we show the observational data (points) of the rotation curve with the corresponding velocity dispersions (error bars) as reported by \cite{Verheijen2001} for NGC 3726, NGC 3877 and NGC 4010. The solid lines correspond to the analytical expressions \eqref{eq6} fitted to the rotation curves. As can be seen, in each case the model fits the observed data with good accuracy. Additionally, in panels (b) of Figures 1, 2, and 3, we plot the normalized mass-density distribution \eqref{eq5} at $z=0$ for the three galaxies, as a function of the dimensionless radial coordinate $\tilde{R}$. Here, we obtain a well-behaved mass-density function, showing a maximum value at the center that decreases to zero at the edge of the disk. On the other hand, in panels (c) of Figures 1, 2, and 3, we present four isodensity curves of the mass-density distribution (\ref{eq5}) in the meridional plane ($\tilde{R}, \tilde{z}$), showing that each model corresponds to a very different mass distribution. Finally, from panels (d) and (e) of the same figures, it is noteworthy that in the three cases the stability conditions are fully satisfied.

\begin{figure}[H] \label{fig3}
\begin{center}
\includegraphics[width=0.459\textwidth]{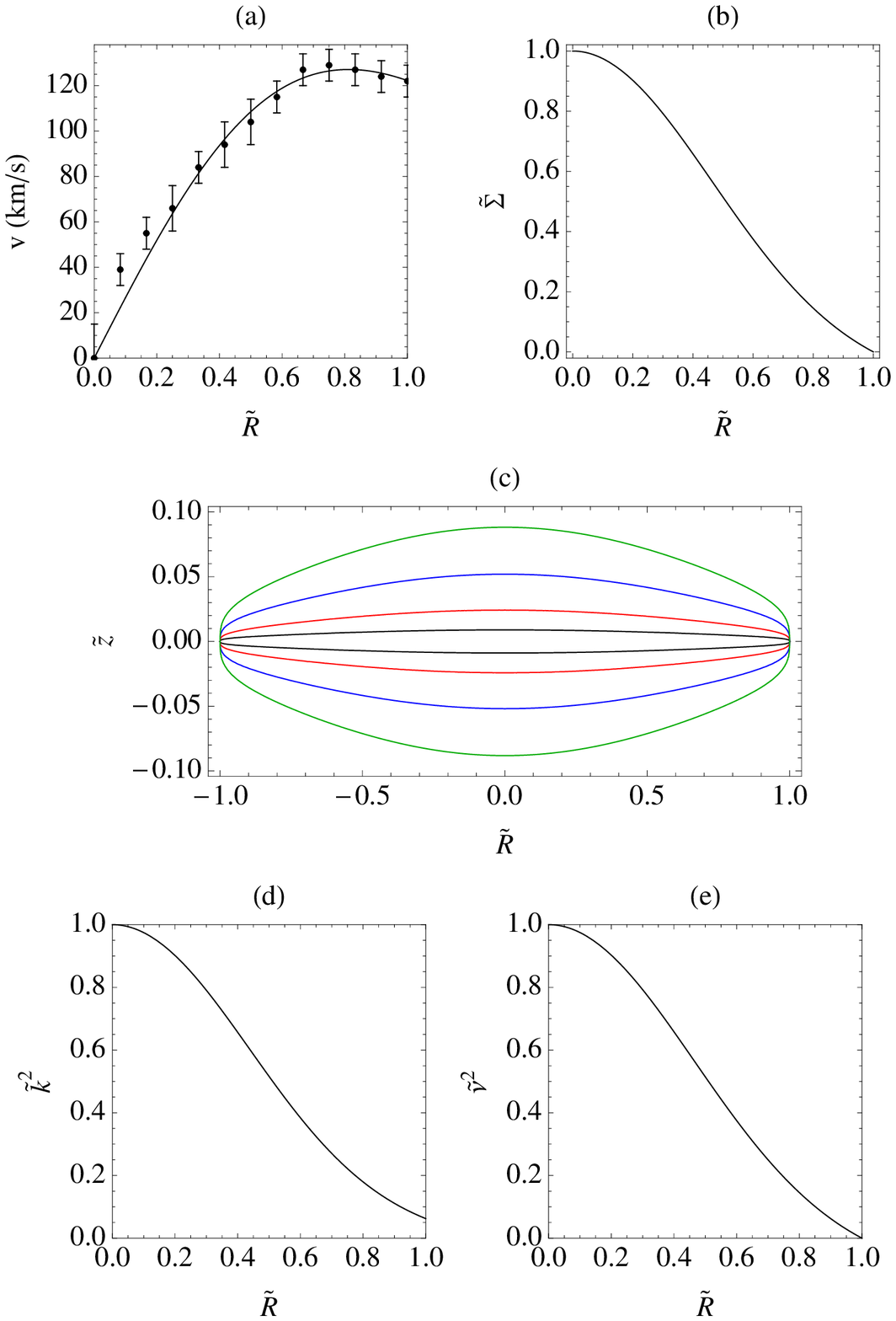}
\end{center}
\end{figure}
\vspace{-8mm}
\begin{small}
{\bf Figure 3.} Model fitted to the rotation curve of NGC 4010 using the parameters given in the third column of Table 1. (a) The solid curve indicates the rotation velocity calculated from \eqref{eq6} while the error bars denote the velocity dispersions of the observational data. (b) Normalized mass-density distribution $\tilde{\Sigma}$ at $z=0$, calculated from \eqref{eq5}. (c) Constant-density curves of equation \eqref{eq5} in the meridional plane. (d)  Epicyclic frequency \eqref{eqEF} evaluated on $z = 0$. (e) Vertical frequency \eqref{eqVF} evaluated on $z = 0$. 
\end{small}

\section*{Stellar Dynamics}
\label{sec:3}

It is a well-known fact that using rough estimates of the dimensions of typical stars and galaxies, the collision interval between stars is about $10^8$ times longer than the average age for most galaxies \citep{Binney2011}. This implies that the star's motion can be determined solely by the gravitational attraction of the galaxy and that collisions between stars are so rare that are irrelevant \citep{Maoz2016}. Therefore, as a first approximation, the orbital dynamics of a star in a given galaxy can be studied following the usual Lagrangian and Hamiltonian approaches for the motion of a test particle in the presence of an estimated gravitational potential. 

The orbital motion of a test particle in an axisymmetric potential is governed by the Lagrangian
\begin{equation}\label{eq:3.1}
\mathcal{L}=\frac{1}{2} \left[\dot{R}^2+(R\dot{\phi})^2+\dot{z}^2\right] - \Phi(R, z),
\end{equation}
with $(R, \phi, z)$ the usual cylindrical coordinates. The generalized canonical momenta read as
\begin{equation}\label{eq:3.2}
p_{R}=\dot{R}, \quad p_{\phi}= R^2\dot{\phi}, \quad p_{z}=\dot{z},
\end{equation}
and the Hamiltonian takes the form
\begin{equation}\label{eq:3.3}
\mathcal{H}=\frac{1}{2}\left(p_{R}^2+p_{z}^2\right)+\Phi_{\rm eff}(R, z),
\end{equation}
with 
\begin{equation}\label{eq:3.4}
\Phi_{\rm eff}(R, z)= \frac{L_{z}^2}{2 R^2}+\Phi(R, z).
\end{equation} 
Here, $ L_{z}=p_{\phi}=$constant, denotes the conserved component of angular momentum about the $z$-axis.

From \eqref{eq:3.3}, the resulting Hamilton's equations of motion can be expressed as 
\begin{eqnarray}\label{eq:3.5}
\dot{R}&=& p_{R}, \\ 
\dot{z}&=& p_{z}, \\ 
\dot{p_{R}}&=&\frac{L_{z}^2}{R^3}-\frac{\partial \Phi(R, z)}{\partial R}, \\ 
\dot{p_{z}}&=&-\frac{\partial \Phi(R, z)}{\partial z},  
\end{eqnarray} 
where $\Phi(R, z)$ is given by Eq. (\ref{eq4}) and its respective parameters should be taken from Table 1.

Since the Hamiltonian is autonomous, $\mathcal{H}$ is an integral of motion 
\begin{equation}\label{eq:3.6}
\mathcal{H}(R,z,p_{R}, p_{z})= \mathcal{H}(R_0,z_0,p_{R_{0}}, p_{z_{0}})=h,
\end{equation} 
with $h$ the energy of an orbit. 

The existence of an analytic integral of motion reduces the phase space dimensionality, and hence the Poincar\'e surface of section is an appropriate and well-established method to analyze the dynamics of the system. Taking into account the axial symmetry associated to the system, it is customary to choose the equatorial plane $z= 0$ as the Poincar\'e plane in order to represent the surface of sections in the $(\tilde{R},\dot{\tilde{R}})$-plane. The orbits were numerically integrated forward in time for 1000 units of time by using a Runge-Kutta-Fehlberg Method (RKF45), with this setting the numerical error related to the conservation of the energy is at most $10^{-14}$. In all cases we set $ z_0 = p_{R_0} = 0$ and we scan the phase space with a large number of initial conditions for the radii $R_{0}$, these three values allow us to determine the values of $p_{z_{0}}$ through the relation \eqref{eq:3.6}.

\begin{figure}[H] \label{fig4}
\begin{center}
\includegraphics[width=0.45\textwidth]{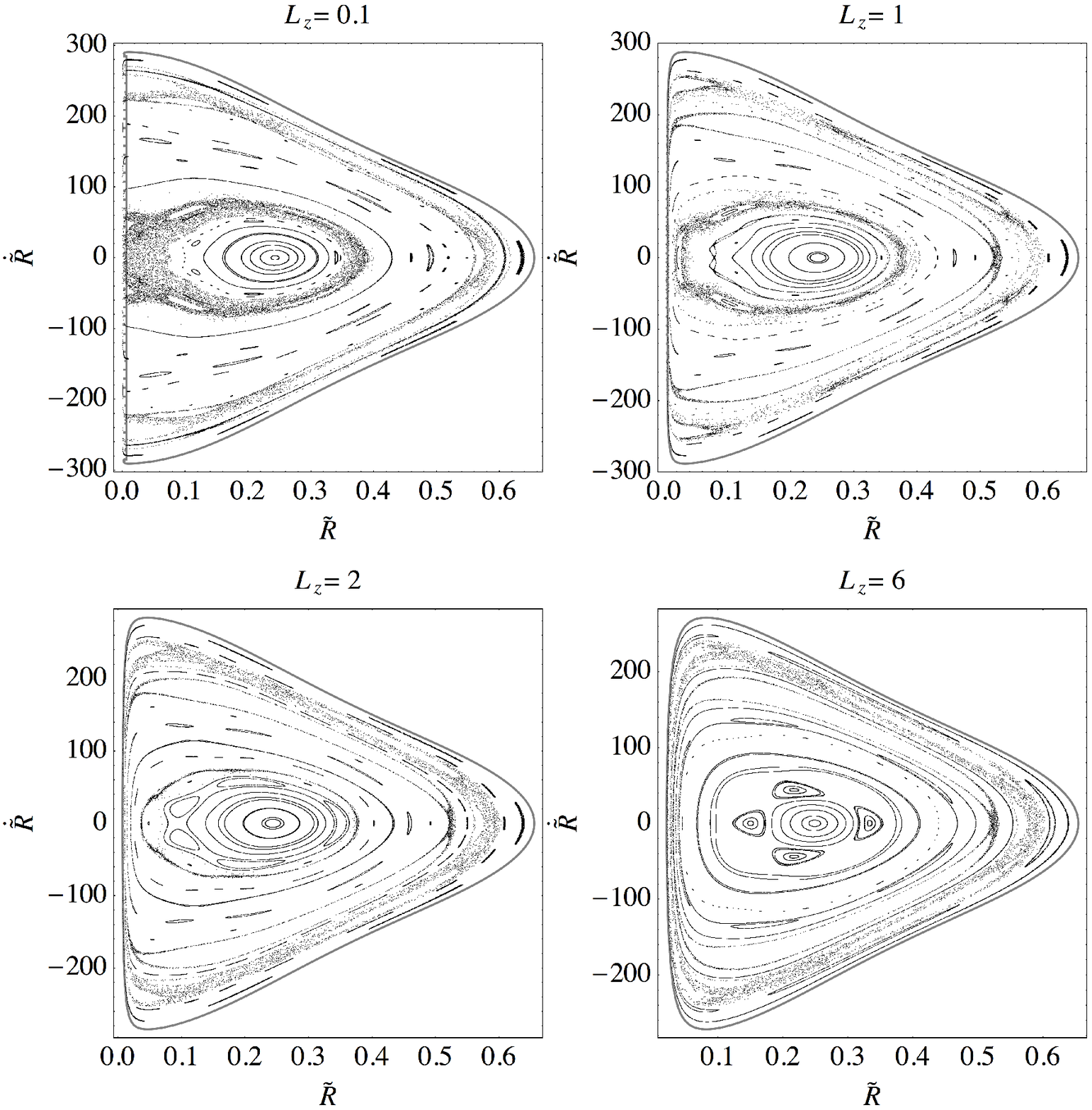}
\end{center}
\end{figure}
\vspace{-8mm}
\begin{small}
{\bf Figure 4.} Poincar\'e surfaces of section of NGC 3726 for different values of angular momentum $L_{z}$ with $h=-1$.
\end{small}

\begin{figure}[H] \label{fig5}
\begin{center}
\includegraphics[width=0.45\textwidth]{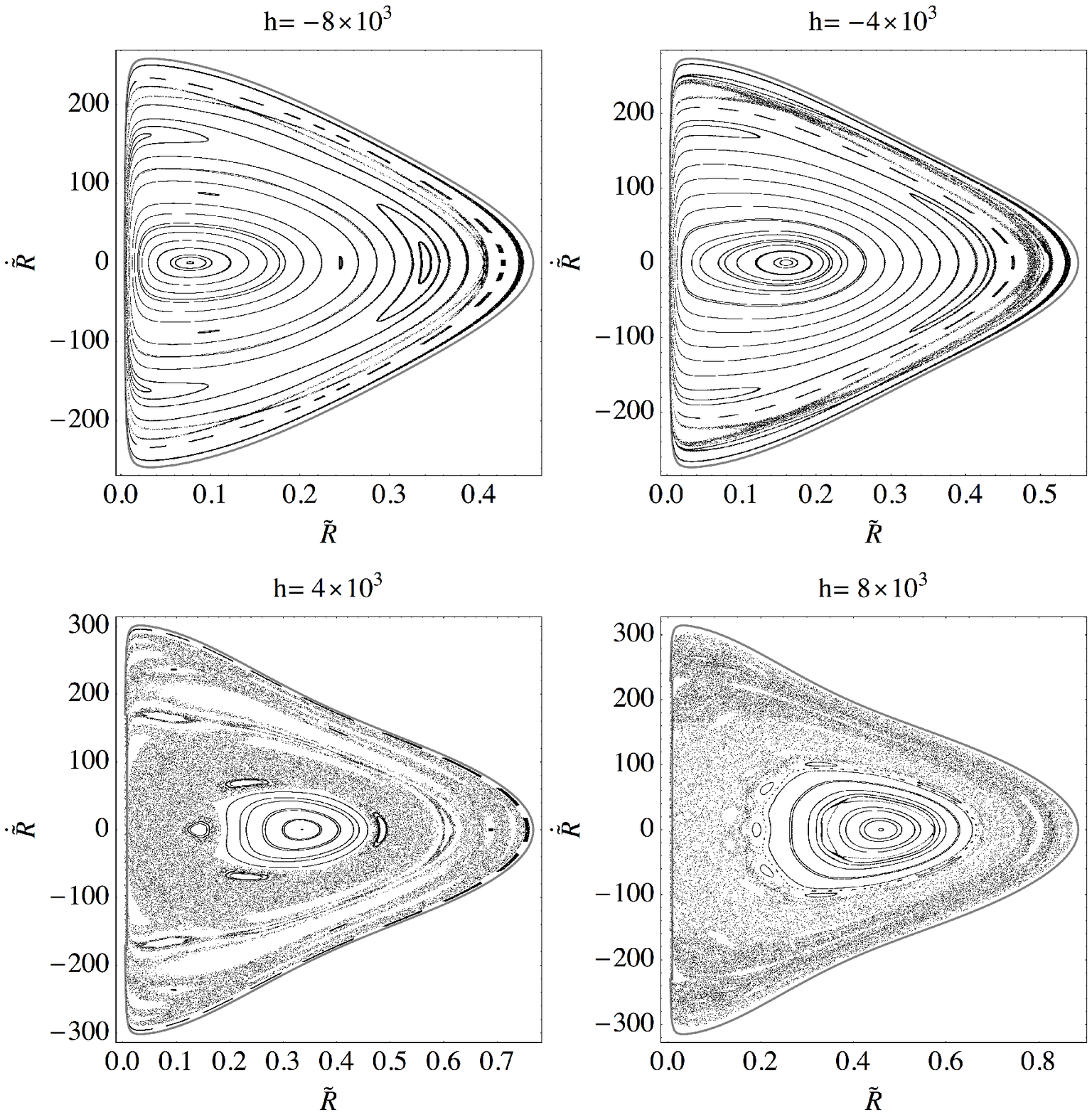}
\end{center}
\end{figure}
\vspace{-8mm}
\begin{small}
{\bf Figure 5.} Poincar\'e surfaces of section of NGC 3726 for different values of total energy $h$ with $L_{z}=1$.
\end{small}

\begin{figure}[H] \label{fig6}
\begin{center}
\includegraphics[width=0.45\textwidth]{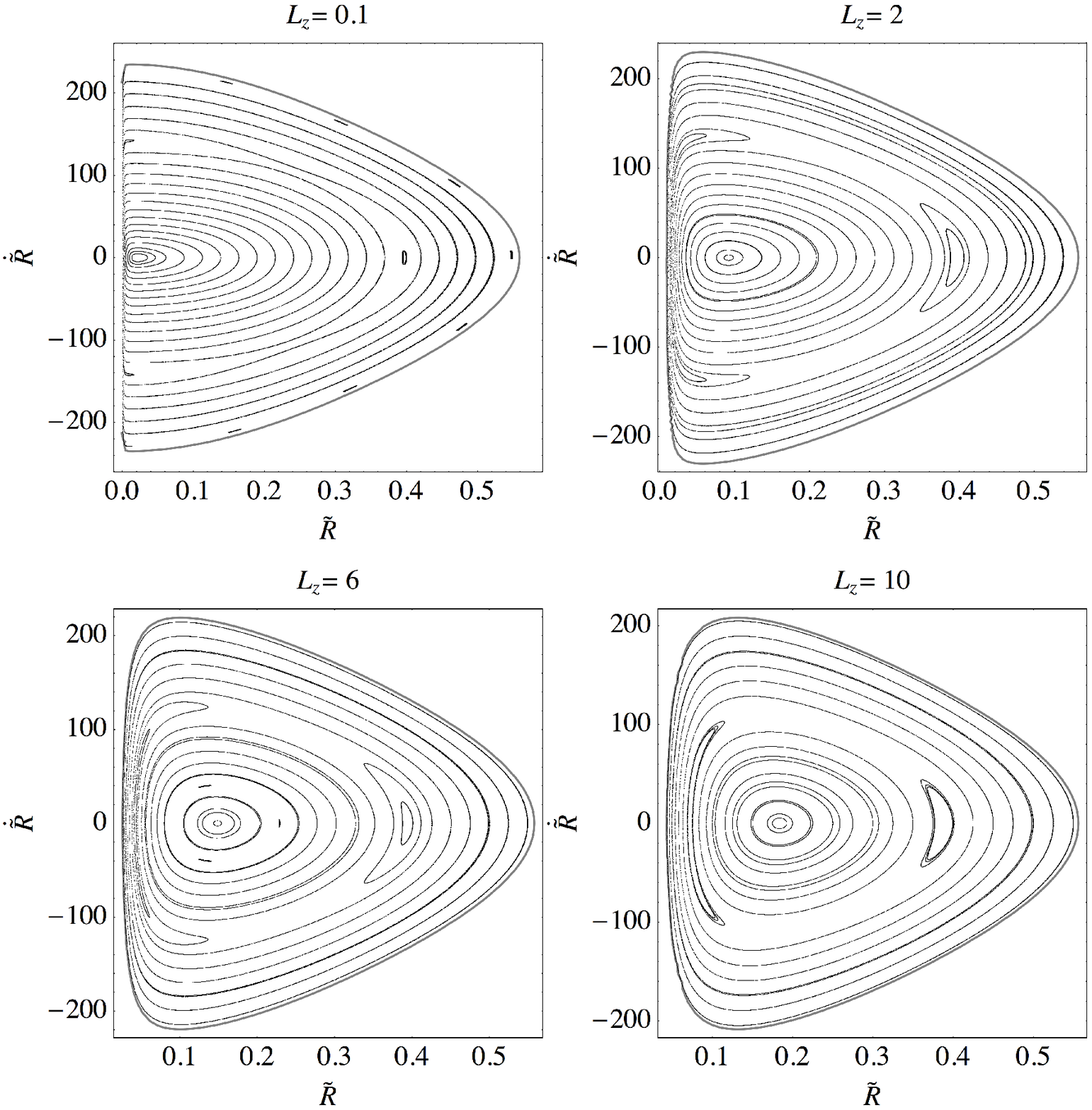}
\end{center}
\end{figure}
\vspace{-8mm}
\begin{small}
{\bf Figure 6.} Poincar\'e surfaces of section of NGC 3877 for different values of angular momentum $L_{z}$
with $h=-1$.
\end{small}

\begin{figure}[H] \label{fig7}
\begin{center}
\includegraphics[width=0.45\textwidth]{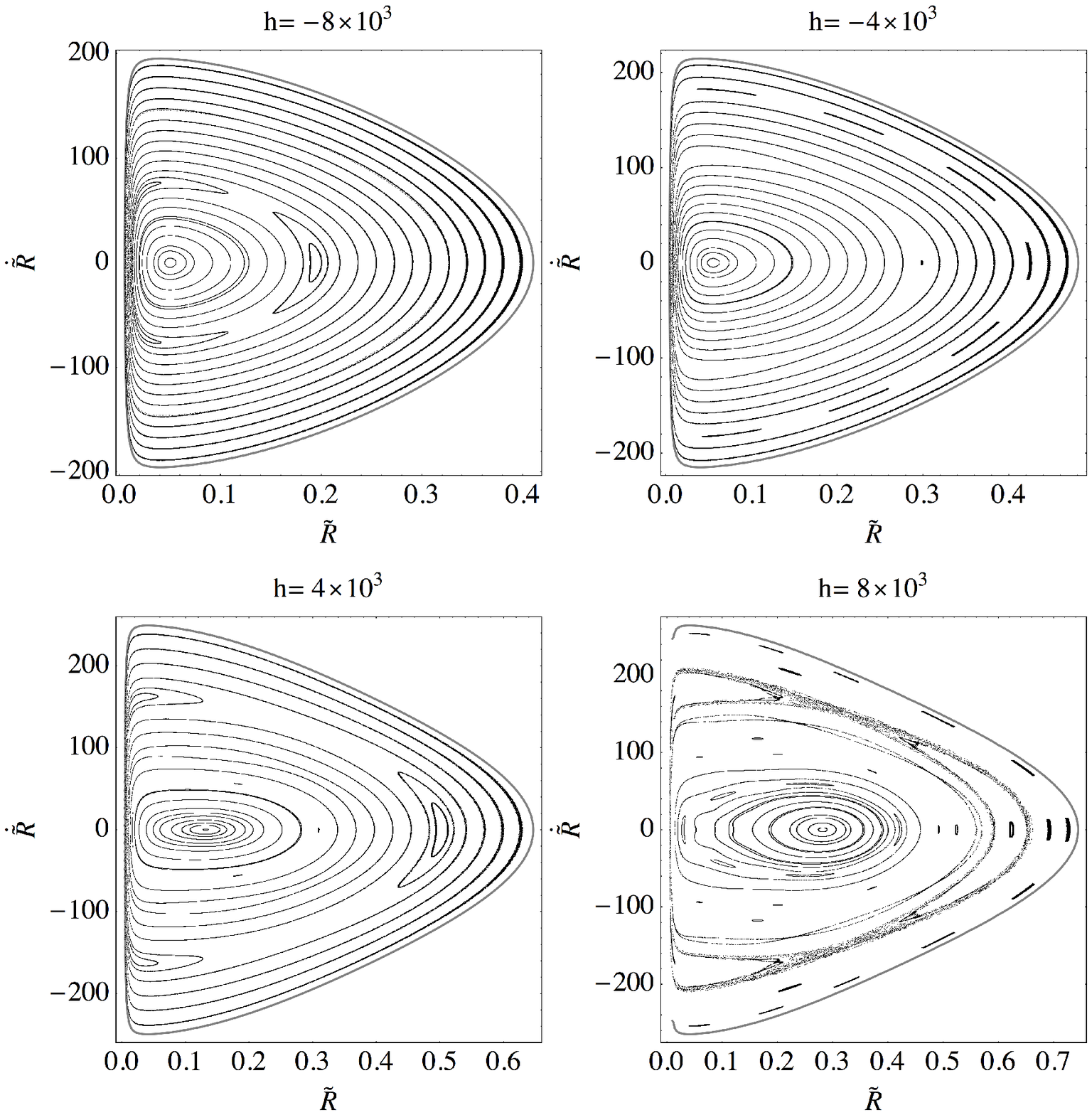}
\end{center}
\end{figure}
\vspace{-8mm}
\begin{small}
{\bf Figure 7.} Poincar\'e surfaces of section of NGC 3877 for different values of total energy $h$ with  $L_{z}=1$.
\end{small}

\begin{figure}[H] \label{fig8}
\begin{center}
\includegraphics[width=0.45\textwidth]{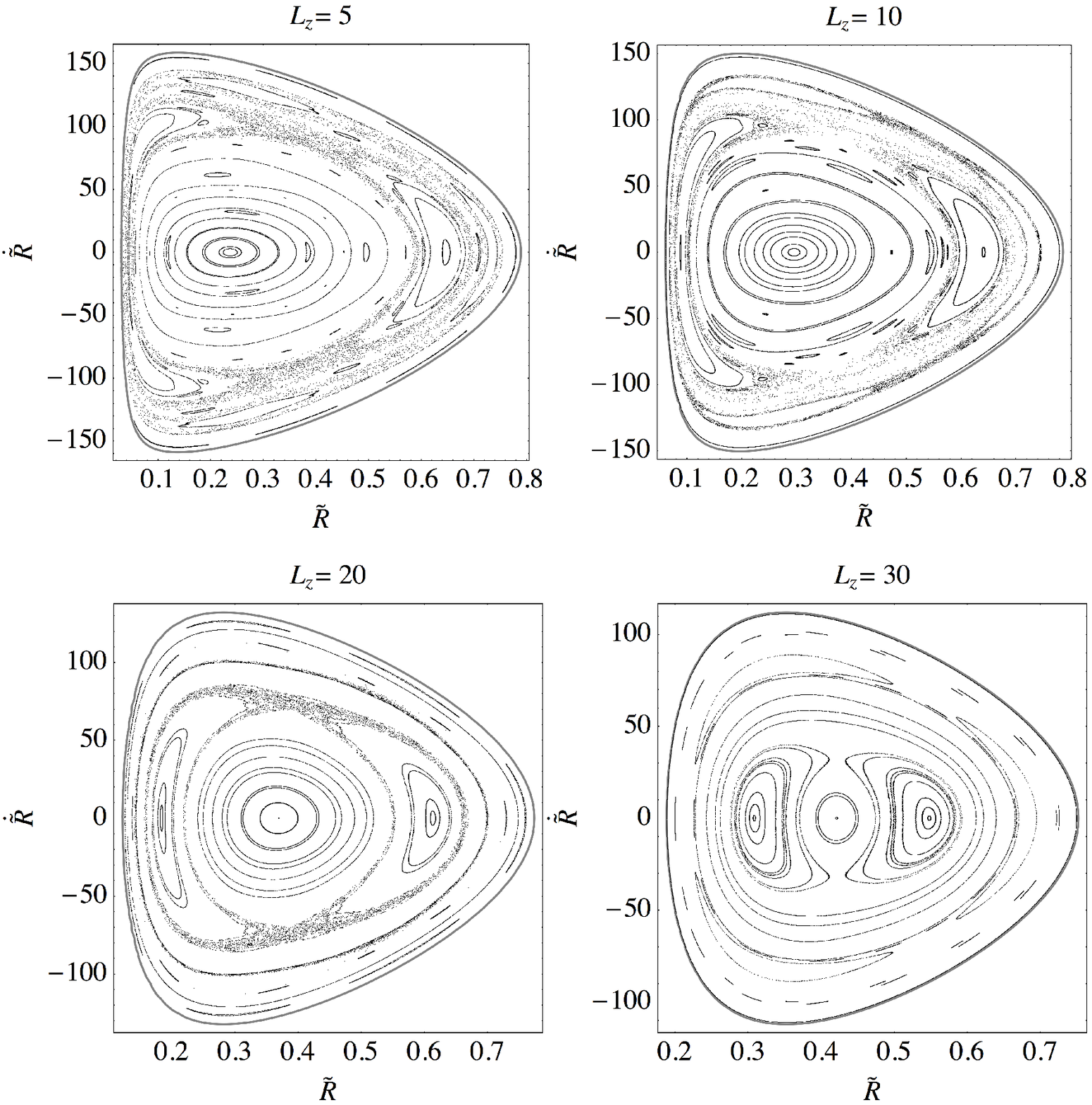}
\end{center}
\end{figure}
\vspace{-8mm}
\begin{small}
{\bf Figure 8.} Poincar\'e surfaces of section of NGC 4010 for different values of angular momentum $L_{z}$ with $h=-10$.
\end{small}

\begin{figure}[H] \label{fig9}
\begin{center}
\includegraphics[width=0.45\textwidth]{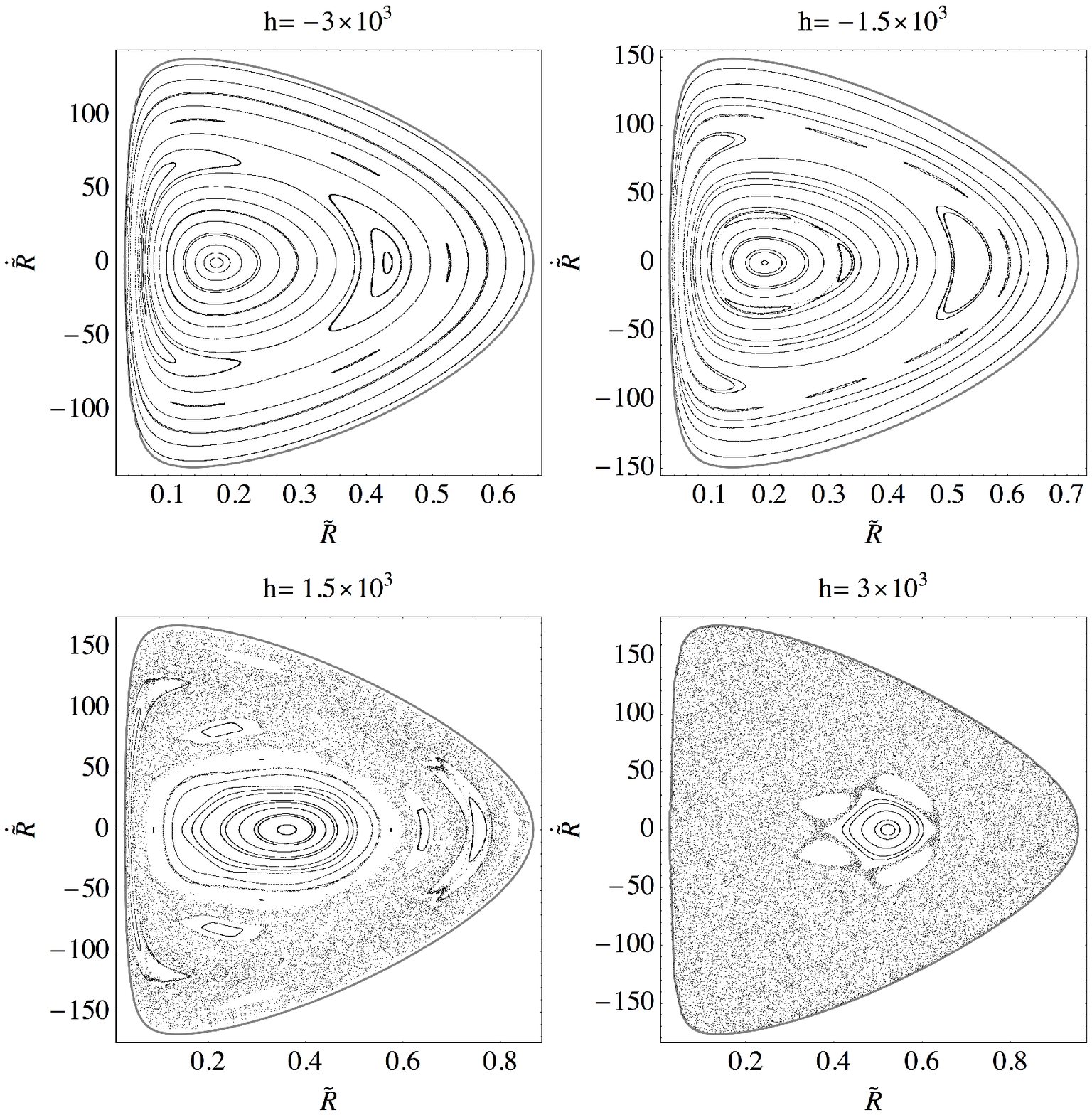}
\end{center}
\end{figure}
\vspace{-8mm}
\begin{small}
{\bf Figure 9.} Poincar\'e surfaces of section of NGC 4010 for different values of total energy $h$ with  $L_{z}=5$.
\end{small}

The transition from regularity to chaos (or viceversa) that takes place for the three considered galaxy models was inspected through the Poincar\'e sections in Figs. 4-9, by using different values of $L_{z}$ (Figs. 4, 6 and 8) and $h$ (Figs. 5, 7 and 9).  It can be observed that the orbital motion exhibits a strong dependence on the angular momentum $L_{z}$ and energy $h$ of the test particle. In particular, from the surfaces of section presented in Figs. 4, 6, and 8, we may infer that there exists an increase in the regularity of the system for larger values of the angular momentum $L_{z}$, i.e. if there exists a chaotic sea the increase of $L_{z}$ will fill the phase with KAM islands, while the opposite effect is observed for larger values of energy $h$ (see Figs. 5, 7, and 9), where the KAM islands deform and shrink giving place to larger regions of chaos.

\section*{Concluding remarks}\label{sec:4}

In the present paper, using the general solution to the Laplace equation, we have derived a generalized Miyamoto-Nagai potential. By means of the nonlinear least square fitting, the analytical velocity curves were adjusted to the observed ones of three specific spiral galaxies: NGC 3726, NGC 3877 and NGC 4010. The resulting analytical models were used to determine the mass-density distributions and the vertical and epicyclic frequencies, showing that unlike the results presented in \cite{Gonzalez2010} for NGC 3877 and NGC 4010, our models satisfy the stability conditions for radial and vertical perturbations. Even though the set of models presented here should be considered as a rough approximation, the circular velocities were shown to fit very accurately to the observed rotation curves and in the three cases the stability conditions were fully satisfied. Here, it is important to note that contrary to the observed tendency in the Miyamoto-Nagay model, where the limit $a\rightarrow 0$ reduces to the Plummer sphere, our models exhibit a tendency to an spherical mass distribution with increasing of the parameter $a$. 

On the other hand, by using the Poincar\'e section method we have also studied the dynamics of the meridional orbits of stars in presence of the gravitational field of the galaxy models. From our results it may be inferred that there exists an increase in the regularity of the orbits for larger values of the angular momentum, while for larger values of energy the orbits tend to be more chaotic. Our toy models suggest that in the three galaxy models chaotic orbits are possible, however the chaotic behavior is very weak for the NGC 3877 model in comparison to NGC 3726 and NGC 4010. It should be noted that none of the studied models showed a fully chaotic phase space. Our results could have significant implications for the study of the dynamics and kinematics of these three specific galaxies, since the regular or chaotic behaviors could shed lights into the evolution and structure of these galaxies, i.e., in phase space, regular orbits are trapped in the vicinity of neighbor orbits, while chaotic orbits, by its own nature, will diverge exponentially in time from its neighbors by filling the phase space in an erratic manner.

\section*{Acknowledgments} 
We would like to thank the anonymous referees for their useful comments and remarks, which improved the clarity and quality of the manuscript. 
FLD, SMM and GAG gratefully acknowledges the financial support provided by COLCIENCIAS (Colombia) under Grants No. 8840 and 8863.

\section*{Authors' contributions} 
All authors make substantial contributions to conception, design, analysis and interpretation of data. All authors participate in drafting the article and reviewed the final manuscript.

\section*{Conflict of interest} 
The authors declare that they have no conflict of interest.

\bibliographystyle{chicago}

\renewcommand{\refname}{\bf References}

\begin{footnotesize}

\end{footnotesize}

\end{multicols}
\end{small}
\end{document}